\documentclass[useAMS]{mn2e}
\usepackage{graphicx}
\usepackage{epsfig}
\usepackage{amssymb}
\usepackage{lscape}
\usepackage{ulem}
\usepackage{txfonts}

\def\kms{km~s$^{-1}$}

\def\c2s{C\,{\sc ii}$^{\star}$}
\def\hkpc{$h_{70}^{-1}$ kpc}

\def\dv{$\Delta$V}

\def\rp{$r_p$}

\title[Radio-loud AGN in mergers] {Galaxy pairs in the Sloan 
Digital Sky Survey - XII: The fuelling mechanism of low excitation radio-loud AGN.}

\author[Ellison et al.] {Sara L. Ellison$^1$,
  David R. Patton$^2$,
  Ryan C. Hickox$^3$\\
$^1$ Department of Physics \& Astronomy, University
of Victoria, Finnerty Road, Victoria, British Columbia, V8P 1A1,
Canada.\\
$^2$ Department of Physics \& Astronomy, Trent University,
1600 West Bank Drive, Peterborough, Ontario, K9J 7B8, Canada.\\
$^3$ Department of Physics \& Astronomy, Dartmouth College, 6127 Wilder 
Laboratory, Hanover, NH 03755, USA.
}
\begin{document}

\maketitle

\begin{abstract}
  We investigate whether 
  the fuelling of low excitation radio galaxies (LERGs) is linked to major galaxy interactions.
  Our study utilizes a sample of 10,800 spectroscopic galaxy pairs and 
  97 post-mergers selected from the Sloan Digital Sky
  Survey with matches to multi-wavelength datasets.
  The LERG fraction amongst interacting galaxies is a factor of 3.5
  higher than that of a control sample matched in local
  galaxy density, redshift and stellar mass.  However, the LERG excess in pairs does not
depend on projected separation and remains elevated out to
  at least 500 \hkpc, suggesting that major mergers are not their main
fuelling channel. In order to identify the primary fuelling mechanism of LERGs, we compile samples of control galaxies
  that are matched in various host galaxy and environmental properties.
  The LERG excess is reduced, but not
  completely removed, when halo mass or D$_{4000}$
  are included in the matching parameters.  However, when \textit{both} $M_{halo}$ and D$_{4000}$ are matched,
  there is no LERG excess and the 1.4 GHz luminosities (which trace jet mechanical power)
  are consistent between the pairs and control.  In contrast, the excess of optical and mid-IR
  selected AGN in galaxy pairs is unchanged
  when the additional matching parameters are implemented.  Our results suggest that whilst major interactions may trigger
  optically and mid-IR selected AGN, the gas which fuels the LERGs has
  two secular origins: one associated with the large scale environment,
  such as accretion from the surrounding medium or minor mergers, 
plus an internal stellar mechanism,
  such as winds from evolved stars.  
\end{abstract}

\begin{keywords}
  Galaxies: interactions,  galaxies: active,  galaxies: Seyfert,
  radio continuum: galaxies
\end{keywords}

\section{Introduction}

The accretion of gas onto a central supermassive black hole results in
the emission of energy over a range of wavelengths, leading
to a variety of available techniques for the identification of active
galactic nuclei (AGN).  Compton up-scattering of UV photons within the
accretion disc leads to X-ray emission, heating of the dust torus is manifest in the
mid-infra-red (mid-IR), broad and/or narrow optical emission lines
may be observed depending on viewing angle and powerful radio jets
emit through synchrotron radiation. Whilst many of these AGN identifiers
are complementary, and different techniques can, in theory, detect a
given AGN, there are also important distinctions between AGN classes.
Importantly, the emerging picture over the last decade is that AGN
can accrete their material in at least
two modes (e.g. Best et al. 2005; Tasse et al. 2008; Hickox et al. 2009; 
Trump et al. 2011; Janssen
et al. 2012; Best \& Heckman 2012; Gurkan, Hardcastle \& Jarvis 2014).  
The first, commonly called `quasar', `radiative' or `high-excitation' mode, refers
to AGN with fairly high accretion rates, in which the material is accreted
from an optically thick, geometrically thin disc.  Such AGN emit their 
energy across a broad 
spectral range, including through optical emission lines, IR radiation 
from the heated dust torus, and, in some cases, radio jets.  The host galaxies
of radiatively efficient AGN tend to be actively star-forming, harbour
relatively low mass black holes and have small stellar bulges (Kauffmann et al. 2003;
Best et al. 2005; Smolcic 2009; Best \& Heckman 2012). However, there may be a minimum
value to the accretion rate that is required in order to yield such radiatively
efficient AGN (e.g. Narayan \& Yi 1995; Trump et al. 2011).  The second type of AGN is therefore radiatively
inefficient (or `low excitation'); these AGN are thought
to be fuelled by cooling, advection dominated accretion flows, or through Bondi accretion,
emitting most of their energy in the form of radio jets, with a characteristic lack
of X-ray, optical line and mid-IR emission (Hine \& Longair 1979; 
Jackson \& Rawlings 1997;  Hardcastle, Evans \& Croston 2006).
The accretion rates of radiatively
inefficient AGN are typically much lower than the high excitation AGN and
have high mass black holes, low rates of star formation and more massive stellar
bulges (Best et al. 2005; Tasse et al. 2008; Kauffmann \& Heckman 2009; Best \&
Heckman 2012).  Therefore, whilst AGN identified at X-ray, mid-IR or
optical wavelengths are generally members of the `high excitation' family,
radio-loud AGN may be either high excitation, or low excitation radio
galaxies (HERGs and LERGs respectively).  The distinction of the HERGs and
the LERGs amongst radio-loud AGN therefore epitomizes the paradigm
of the two fuelling modes of AGN accretion (Best et al. 2005; Hardcastle,
Evans \& Croston 2007; Best \& Heckman 2012).

Underlying the existence of the two AGN accretion modes is the fundamental question of
what physical mechanisms are responsible for the delivery of gas to the galactic centre,
and whence this gas originates.  It has been suggested that high excitation
(radiatively efficient) AGN require a plentiful supply of cold gas, such as
could be supplied during a merger (e.g. Kauffmann \& Heckman 2009; Johansson,
Naab \& Burkert 2009; Capelo et al. 2015).  Indeed, there are now many studies
that have shown that interactions over a wide range of mass
ratios \textit{can}\footnote{This is not to say that
  all radiatively efficient AGN are triggered by mergers; both observations
  and simulations support the likelihood that other mechanisms, such as
  cosmic inflows or internal instabilities could also contribute, or even dominate
  (Draper \& Ballantyne 2012; Menci et al. 2014).}
trigger AGN that are selected variously in the optical
(e.g. Alonso et al. 2007; 
Ellison et al. 2011, 2013;  Kaviraj 2014; Khabiboulline et al. 2014), mid-IR
(Satyapal et al. 2014; Shao et al. 2015) and X-ray (Koss et al. 2010, 2012;
Silverman et al. 2011; Lackner et al. 2014).
For radio selected AGN, there is evidence to support a connection between
mergers and both HERGs and the highest luminosity radio-loud AGN
(e.g. Ramos-Almeida et al. 2011; Tadhunter et al. 2011; Kaviraj et al. 2015).

In contrast to the radiatiatively efficient AGN,
theoretical models, and their comparison to observations, indicate that the radiatively
inefficient accretion that leads to LERGs need not be interaction-induced,
and can be maintained through
low level accretion from hot gas in galactic halos, either through a Bondi-like
process, or through cooling
(Best et al. 2005, 2006; Allen et al. 2006;  Hardcastle et al. 2007; Gaspari,
Ruszkowski \& Oh 2013).  Indeed,
elliptical galaxies are known to be surrounded
by large reservoirs of hot gas, created by mass loss from evolved
stellar populations and fed from external reservoirs such as the
intergalactic and intra-cluster media (IGM and ICM respectively, see
Mathews \& Brighenti 2003 for a review) or through minor mergers.
The dependence of LERG prevalence on environment indicates that some of
the gas that fuels the AGN may indeed originate in the IGM or ICM
(e.g. Best et al. 2007).

Although the picture of stochastic, low accretion rates in LERGs is generally
considered to operate independently of mergers,
several recent works have presented observations that indicate
that interactions \textit{do} contribute to the fuelling of LERGs.  For example,
Sabater et al. (2013) found that the fraction of LERGs is enhanced by interactions,
as inferred through a tidal force indicator.  The LERG excess is detected
even when the stellar mass of the galaxy (correlated with the galactic halo mass,
and hence an indirect measure of the hot halo reservoir) is accounted for.
Pace \& Salim (2014) find
an excess of close companions (within 100 kpc) around LERGs compared to
radio-quiet galaxies.  The excess of satellites persists even when
cluster membership is controlled for, indicating that the availability of
hot gas (either internal or external) is not the only requirement
for producing a LERG. These and other works  (e.g. Tasse et al. 2008)
have concluded that interactions with near neighbours may
contribute to the fuelling of LERGs.  
In this paper, we aim to further test whether major interactions,
between approximately equal mass galaxies, play a role in the fuelling of LERGs,
as they apparently do for
the high excitation AGN selected in the optical, mid-IR and X-ray.

\section{Samples and derived properties}\label{sample_sec}

The samples of galaxy pairs and post-mergers used in this work have been described
fully in the previous papers in this series.  We provide here brief
details and refer the reader to our earlier work for more information
on parent samples, effects of mass ratio, bias due to fibre collisions
and morphological classifications
(e.g. Ellison et al. 2008, 2010; Scudder et al. 2012; Patton et al. 2013).

We adopt the sample of spectroscopically identified pairs presented in
Ellison et al. (2013).  In brief, the pairs are selected from
the Sloan Digital Sky Survey Data Release 7 (SDSS DR7) 
Main Galaxy Sample (14.0 $\le m_r \le$ 17.77) with a
redshift range $0.01 \le z \le 0.2$ and SDSS specclass=2.  The
pairs are required to have a projected separation \rp\ $\le$ 80 \hkpc,
a velocity difference \dv\ $\le$ 300 \kms\ and a stellar mass ratio
within a factor of 4 of one another\footnote{Although the $\Delta V < 300$
\kms criterion mitigates projection effects, it is likely that not
all of the pairs will result in a final merger.  E.g. Moreno et al. (2013)
find that as few as $\sim$ 10 per cent of wide separation (\rp$<$ 250 \hkpc)
pairs are dominantly bound to one another.}.  After rejecting 
67.5 per cent of pairs with angular separations $\theta > 55$
arcsec to account for fibre collision selection effects, there are 10,800
galaxies in the pairs sample. Stellar masses, photometric properties,
4000 \AA\ break strengths ($D_{4000}$) and total SFR
are taken from Mendel et al. (2014), Simard et al. (2011)
and the JHU/MPA catalogs (e.g. Brinchmann et al. 2004), respectively.

In addition to the sample of close pairs, which probes the pre-coalescence
phase of the merger, we also include in the analysis a sample of
post-merger galaxies.  The post-mergers represent visually selected galaxies
that exhibit strong morphological disturbances and are likely the remnants
of major mergers.  The post-merger sample is described in detail in
Ellison et al. (2013; 2015), in which it is shown that the post-coalescence
phase corresponds to the highest SFR enhancements and AGN frequency,
compared to matched control samples.  After careful visual classification
and the requirement of stellar mass availability (see below), there are 97 galaxies
in the post-merger sample.

Three definitions of AGN are used in this work, which, for brevity, we
refer to as optical, mid-IR and radio-selected.  Following Ellison
et al. (2013), the optical AGN are selected according to the emission
line criteria of Stasinska et al. (2006), which identifies galaxies with
even a small contribution from AGN\footnote{Although the exact number of
  AGN depends on the optical emission line threshold adopted, our
  results are independent of selection technique, e.g. see Ellison
  et al. (2011) for a comparison between optical AGN selection
  methods.}.  For the mid-IR selected AGN, we follow the work of
Satyapal et al. (2014) and use a Wide Field Infra-red Explorer (WISE)
W1$-$W2$>$0.5 colour cut to select dust obscured AGN\footnote{Although
  W1$-$W2$>$0.5 is more liberal than the threshold of 0.8 that is
  sometimes used, its applicability at low redshift is demonstrated
  in Satyapal et al. (2014).  Adopting a more stringent cut does not
affect the conclusions of our work.}.  The radio-selected
AGN are taken from the compilation of Best \& Heckman (2012), who match
the SDSS to NVSS and FIRST catalogs, divide radio detections into star-forming
or AGN categories and use a combination of criteria to distinguish the
latter into HERGs and LERGs.  In this work, we select only the LERGs to constitute
the sample of radio-loud AGN.  The LERGs in our sample trace relatively low
luminosity AGN, with most galaxies in the range $23< \log L_{NVSS, 1.4 GHz} < 25$
W Hz$^{-1}$ (see also Best \& Heckman 2012).

\section{Methodology and matching parameters}\label{methods_sec}

In previous papers of this series, we have employed a matching
methodology between the galaxy pairs and a control sample,
in order to identify changes induced by the interaction.
Our standard procedure has been to match each galaxy in a
pair with a sample of control (no neighbour within 80 \hkpc\ and 10,000
\kms) galaxies of the
same stellar mass, redshift and local density. The
latter parameter is quantified using $\Sigma_n = \frac{n}{\pi d_n^2}$,
where $d_n$ is the projected distance in Mpc to the $n^{th}$ nearest
neighbour within $\pm$1000 \kms.  We adopt $n=5$ for this work, but note that
our results do not depend sensitively on the choice of $n$. 
Normalized densities, $\delta_5$,
are computed relative to the median $\Sigma_5$ within a redshift slice
$\pm$ 0.01.  For each galaxy in a pair/post-merger, all possible control galaxies
are identified within some specified tolerance.  The tolerance for matching
is 0.005 in redshift, 0.1 dex in stellar mass and 0.1 dex in normalized
local density.  If less than five matches are found, the
tolerance is grown by a further $\Delta z$=0.005 in redshift,
$\Delta$log M$_{\star}$=0.1 dex in stellar mass and  $\Delta \delta_5$=0.1 dex
in normalized local density until the required number of matches is achieved. 
For each galaxy in a pair/post-merger, we can hence compare the average properties
of its matched controls with the interaction/merger itself.  Using this technique, we have
previously reported trends between the projected separation of galaxy pairs
and changes in SFR (Patton et al. 2013), gas-phase metallicity (Scudder et al. 2012),
optical and mid-IR AGN fractions (Ellison et al. 2011, 2013; Satyapal et al. 2014)
and black hole accretion rate (Ellison et al. 2013).  The dependence of these
quantities on projected separation is a strong indication that the interaction is
responsible for changes therein.

The fuelling of the LERG AGN class has been variously suggested to be linked
to the presence of close companions, accretion from the old stellar populations
in the bulge, or gas accreting from the hot halo (e.g. Hopkins \& Hernquist
2006; Kauffmann \& Heckman 2009;
Ramos-Almeida et al. 2013; Sabater et al. 2013; Karouzos et al. 2014; Pace \& Salim 2014).  In the latter scenario,
the source of gas feeding the hot halo may either be internal (from the winds of
evolved stars, or supernovae), or may be fed from the external IGM or ICM
(Mathews \& Brighenti 2003).  In order to distinguish between these possibilities,
we perform various combinations of matched parameters.  Stellar mass alone is
likely to be an indicator of the material in the galactic halo with an internal origin.
$\delta_5$ traces local density, whilst the group halo mass (which we adopt from
the DR7 catalog originally published by Yang, Mo \& van den Bosch 2009) informs us about the
reservoir of gas that may be accreted from within a given dark matter halo.
Stellar mass, $\delta_5$ and halo mass can therefore all be considered as
tracers of environment on different scales.

\begin{figure}
\centerline{\rotatebox{0}{\resizebox{8cm}{!}
{\includegraphics{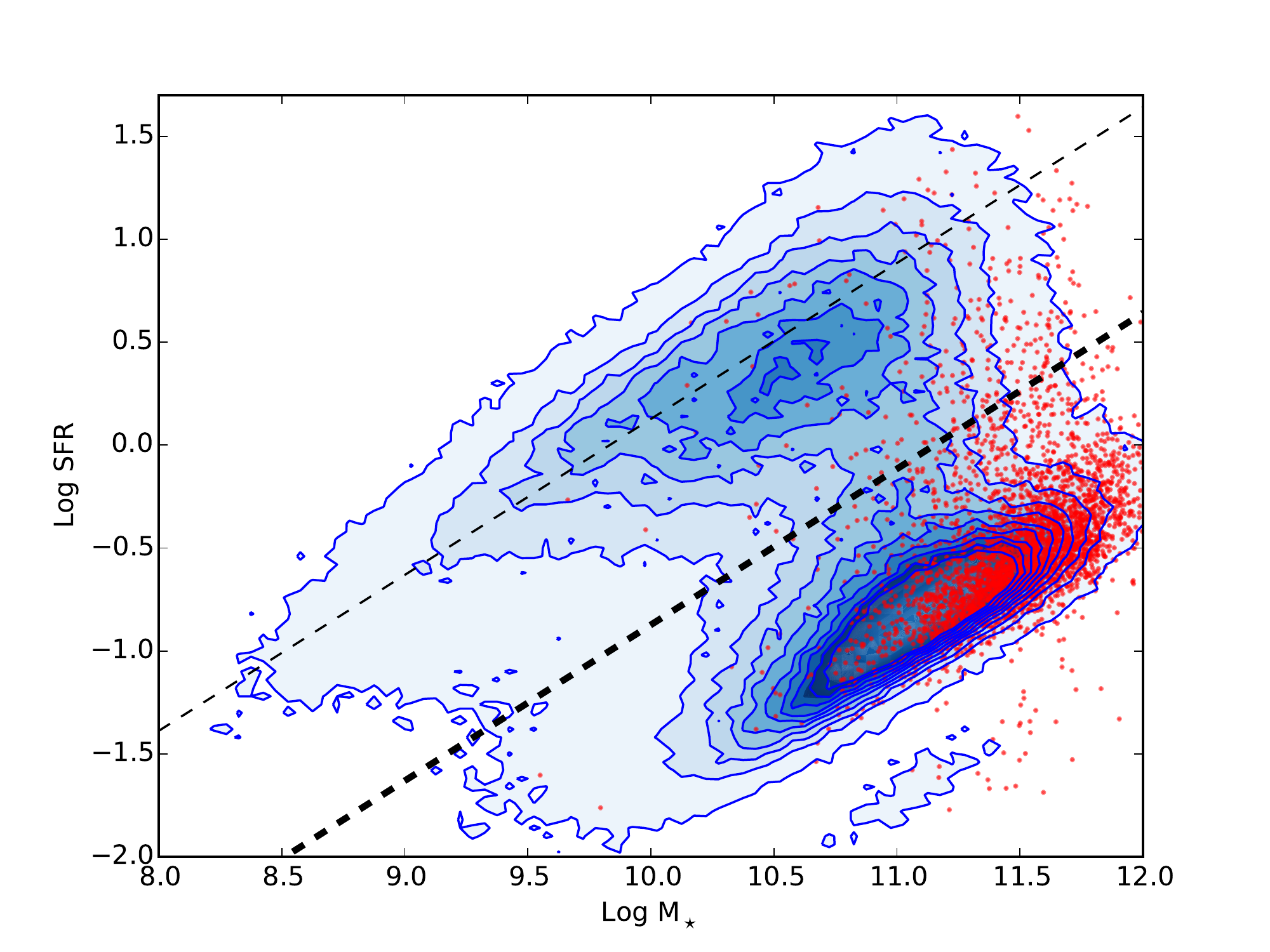}}}}
\caption{\label{ms}The star formation rates and stellar masses of SDSS galaxies
  are shown as contours; two sequences are clearly visible. The star-forming sequence
  is fit with a first order linear least squares fit, as shown by the light dashed
  line.  The fit parameters are log SFR =  -7.4485 + log M$_{\star} \times$0.7575.
  The heavy dashed line is translated downwards by 1 dex in log SFR at fixed
  stellar mass and divides our definition of the star-forming and passive sequences.
  Points show LERGs selected from Best \& Heckman (2012).}
\end{figure}

In addition to environmental parameters, we consider variables that
reflect the star formation properties of
the galaxy.  Best \& Heckman (2012) have shown that LERGs are preferentially passive,
lying below the star-forming main sequence.    By performing a simple first order linear least
squares fit between stellar mass and SFR for star forming galaxies in the SDSS,
and translating the fit downwards by a factor of 10 in SFR, we define the
threshold between the star-forming and passive sequences.  This is illustrated
in Fig. \ref{ms} where the contours show all SDSS galaxies, the dashed lines
show the fit to the star-forming main sequence (iclass=1 from Brinchmann et al.
2004) and its downward translation.
The points show the position of the LERGs identified by Best \& Heckman
(2012); as shown in that paper, the LERGs are preferentially located on the
passive sequence, at relatively high stellar masses.

The crudest way to match
star forming properties is to require that all controls matched
to a given galaxy pair/post-merger are on the same `sequence', i.e. either the passive,
or the star-forming sequence.  We can also attempt to match more precisely
in SFR.  However,  whilst we can be fairly
confident that a given galaxy is on the star-forming or passive sequence,
the exact SFRs of galaxies that are quenched are quite uncertain.  At least
part of this uncertainty is associated with the determination of SFRs from
either low S/N emission lines, or with the calibration from $D_{4000}$ (Brinchmann et al. 2004).
To try to mitigate these problems, we match on the observed  $D_{4000}$,
rather than the inferred SFR.  Matching on $D_{4000}$ is also appropriate given that we are
interested in the link between the mass loss of the old stellar population
and LERG fuelling, so that the instantaneous SFR is less relevent than
the integrated value traced by $D_{4000}$ (although the two properties are correlated).  
Finally, we investigate matching
on the stellar mass of the bulge since Kauffmann \& Heckman (2009) have
identifed a regime of black hole growth that is proportional to the bulge
mass, again implicating fuelling by mass loss from old stars.

\section{The fuelling mechanisms of AGN}\label{results_sec}

\subsection{LERG fractions in major galaxy mergers}

In Table \ref{excess_tab} we present a selection of combinations
of matching parameters that were investigated in our quest to
identify the source of gas that accretes to form LERGs.  For convenience,
we have labelled the sets of matching parameters that we will discuss
in this work.  In our
previous work, our basic matching scheme has used redshift,
stellar mass and $\delta_5$, which we will use as our reference
parameters in this work (Set \#1).

\begin{table}
\begin{center}
\caption{Average excess of LERG AGN in galaxy pairs}
\begin{tabular}{llr}
\hline
Set & Matched parameters & Average pair\\
\# & & LERG excess \\
 \hline	
1 & z, M$_{\star}$, $\delta_5$ & $\times 3.5\pm0.3$ \\
2 & z, M$_{\star}$ & $\times 3.8\pm0.4$ \\
3 & z, M$_{\star}$, $\delta_5$, M$_{halo}$ &  $\times 2.0\pm0.2$\\
4 & z, M$_{\star}$, $\delta_5$, M$_{bulge}$ &  $\times 2.3\pm0.2$\\
5 & z, M$_{\star}$, $\delta_5$, sequence & $\times 2.8\pm0.3$ \\
6 & z, M$_{\star}$, $\delta_5$, D$_{4000}$ &  $\times 2.0\pm0.2$\\
7 & z, M$_{\star}$, $\delta_5$, D$_{4000}$, M$_{halo}$ &  $\times 1.1\pm0.1$\\
8 & z, M$_{\star}$, D$_{4000}$, M$_{halo}$ &  $\times 2.1\pm0.4$\\
\hline
\end{tabular}
\label{excess_tab}
\end{center}
\medskip
\end{table}

\begin{figure}
\centerline{\rotatebox{0}{\resizebox{10cm}{!}
{\includegraphics{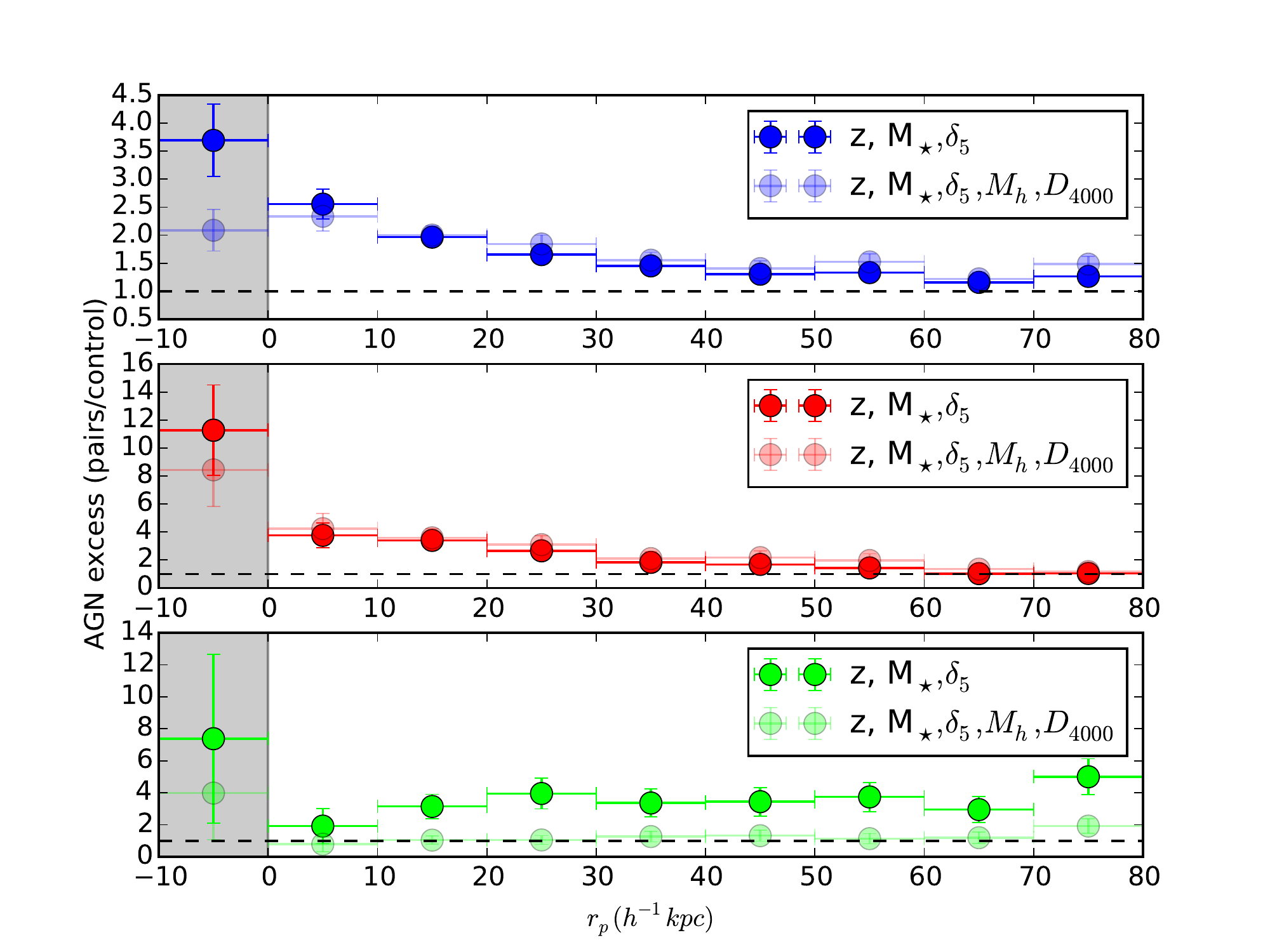}}}}
\caption{\label{mhd4} The fraction of AGN in galaxy pairs and post-mergers (grey box)
  relative to their matched
  controls (the AGN excess) as a function of projected separation.  Optically selected,
  mid-IR selected and radio selected are shown in the top, middle and bottom
  panels respectively.  In all panels, the bright colours show the fiducial matching
  set (Table \ref{excess_tab} Set \#1) and the pale colours represent Set \#7.
  The optical and mid-IR selected AGN show a trend of increasing AGN excess towards
  smaller projected separations; this trend persists even when D$_{4000}$ and halo
  mass are matched.  Although there is an excess of LERGs (lower panel) for the
  fiducial matching set, there is no trend with projected separation, and the excess
disappears when  D$_{4000}$ and halo mass are matched.}
\end{figure}

For the varying combinations of matching parameters we calculate the
AGN excess, defined as the fraction of galaxies classified as AGN in
the mergers (pairs or post-mergers) relative to the fraction of AGN
in their matched controls. We remind the reader that our sample
selects likely \textit{major} mergers, based on the mass ratio
of the pairs (within 4:1) and the morphology of the post-mergers (see Ellison
et al. 2013). In Fig. \ref{mhd4} we show the AGN excess
in the optical (top panel), mid-IR (middle panel) and LERG (bottom panel) AGN for our fiducial
Set \# 1.  The optical
and mid-IR results have been previously published in Ellison et al. (2013)
and Satyapal et al. (2014), but are presented here for comparison
with the radio-selected LERGs.  Fig. \ref{mhd4} shows that there
is an average excess of LERGs in close pairs by a factor of
3.5, in qualitative agreement
with the results of Sabater et al. (2013) and Pace \& Salim (2014).
The LERG excess is even higher in the post-mergers: a factor of $\sim$ 8.
However, unlike the optical and mid-IR AGN fractions, there is no
dependence of the LERG excess on projected separation.  Moreover, by using a
sample of wide separation pairs (Patton et al. 2013) and adopting
the control procedures described in Patton et al. (in prep) we
find that the excess of LERGs in pairs continues out to separations of
at least 500 kpc.  Interactions
are unlikely to precipitate changes in galaxy properties over
such large separations.  For example, star formation rate enhancements
in the same wide pairs sample persist only out to $\sim$ 150 kpc
(Patton et al. 2013).  It is therefore
possible that the LERG excess in the lower panel of Fig. \ref{mhd4}
is not due to interactions, but some
residual dependence of the LERG fraction on other galaxy properties.

We quantify the AGN excess for the various combinations of matching
parameters listed in Table \ref{excess_tab}.  We begin by considering
environmental metrics, since radio-loud AGN are known to
be located in preferentially over-dense regions.  Local density ($\delta_5$)
apparently has little impact on the LERG excess, since removing it
from the matching parameters (Set \#2)
does not significantly change the average LERG excess.  Adding halo mass (matched
to within a tolerance of 0.1 dex) to
the fiducial matching set (Set \#3) does reduce the LERG excess, but there
remains a factor of two more LERGs in pairs than in their
control.  Environment on the group scale therefore seems to
modulate the LERG fraction to some extent, but is apparently not
fully responsible for the fuelling (consistent with the conclusions
of Sabater et al. 2013 and Pace \& Salim 2014).

We now turn to parameters that match the properties of the
stellar component of the merging galaxy itself.   Adding the
stellar mass of the bulge (Mendel et al. 2014, matched to within a tolerance
of 0.1 dex) to our fiducial parameter set (Set \#4)
slightly reduces the LERG excess, but there remains 2.3
times more LERGs in pairs than their controls.   As shown in
Fig. \ref{ms}, LERGs are preferentially located in passive galaxies;
without accounting for this preference many controls (e.g. in Set \# 1)
will be drawn from the star-forming sequence.  This will lead
to an enhanced LERG excess in pairs relative to the control sample.
A modest reduction in the LERG excess (relative to Set \# 1)
is seen when requiring that control galaxies
are on the same `sequence' (passive or star-forming, Set \#5) as the
pairs/post-mergers.  Matching on D$_{4000}$ (to within a tolerance of 0.05, Set \#6) performs slightly better,
reducing the LERG excess to a factor of two.  The result is unchanged
if we restrict the sample to only include galaxies in which the
fibre is dominated by light from the bulge (based on bulge-disc decompositions
from Simard et al. 2011).  Therefore,
as was the case for environmental properties, matching
on parameters associated with the star forming properties
of the galaxy does seem to play a role in LERG fuelling,
but is not the full story.  

We now add both halo mass and D$_{4000}$ to
our fiducial set of matching parameters (Set \#7).  We are thus
accounting for both internal stellar properties and
large scale environment.  The results are shown in pale
coloured points in Fig. \ref{mhd4} for optical, mid-IR and
radio-selected AGN.  Interestingly, the optical and mid-IR
AGN excess is little affected by the additional matching,
and the increasing AGN excess with decreasing projected
separation persists.  Only the post-mergers show a significant
decrease in their AGN excess.
However, in the lower panel of Fig. \ref{mhd4} we show that the 
LERGs respond very differently to the additional matching
of halo mass and D$_{4000}$.  There is now no significant excess of
radio-loud AGN in the pairs (and a statistically insignificant
excess in the post-mergers).  \textit{From this we conclude that the gas that
fuels the LERGs can originate from either external origins, or from
stellar sources within the galaxy.  Once these two factors have been
controlled for, there is no evidence that major galaxy interactions provide an
additional fuelling mechanism.}    Set \#7 includes two parameters
that are linked to `environment', $\delta_5$ and halo mass, and both
are apparently necessary to fully account for the LERG fuelling mechanism\footnote{Whilst
there is a broad correlation between $\delta_5$ and halo mass, a large scatter
exists.}.
This is demonstrated by removing $\delta_5$ in Set \#8, which yields
an average pair excess of $\sim$ 2.  We speculate that the dual contribution from halo mass
and $\delta_5$ exists because whilst the former
is indicative of the total reservoir of halo material, $\delta_5$
measures the local scale distribution of this material.  Alternatively,
the two parameters may trace different sources of external accretion,
such as minor mergers and `smooth' IGM accretion.

We also investigate whether the
mechanical energy associated with the radio jet is enhanced in
the pairs, compared to the control.  Following Best \& Heckman (2012)
we compute $L_{mech} = 7.3 \times 10^{36} (L_{1.4GHz, NVSS}/10^{24} WHz^{-1})^{0.70}$ W,
and compare $L_{mech}$ in each paired galaxy with the median $L_{mech}$ of its
matched controls to determine $\Delta L_{mech} = \log L_{mech,pair} - \log L_{mech,control}$.  
The median $\Delta L_{mech}$ = 0.002, indicating that there is no significant
enhancement in the jet mechanical power of LERGs in close galaxy pairs.
This contrasts with the accretion luminosities derived from both optical and
mid-IR selected AGN, which are enhanced in the close pairs sample (Ellison et al. 2013;
Satyapal et al. 2014).

\section{Conclusions}

Based on a sample of close galaxy pairs and recently coalesced post-mergers
selected from the SDSS, we have investigated the origin of nuclear fuelling
in low excitation radio galaxies.  The frequency of LERGs is a factor of
3.5 higher in the pairs and a factor of 8 higher in post-mergers
than in our fiducial control sample
that is matched in stellar mass, redshift and $\delta_5$.  However,
unlike the excess of optical and mid-IR selected AGN identified in the
same sample, there is no dependence of LERG excess on pair separation,
and the excess persists out to at least 500 \hkpc, indicating that
major interactions are not the cause of the nuclear activity.  Although the
LERG excess is mildly reduced when either bulge mass, D$_{4000}$ or
halo mass are controlled for, LERGs are still a factor of at least
2 higher in the pairs than the control.  However, when we simultaneously
add both halo mass and D$_{4000}$ to the fiducial control sample parameters,
the LERG excess disappears.  There is also no enhancement in the average
mechanical jet luminosity in the pairs, relative to the control.
Our results indicate that the gas responsible for fuelling low excitation radio loud
AGN is not supplied by major mergers, but rather has
has two contributing secular sources: internal stellar processes
(such as winds from old stars) and external accretion, which could include
both minor mergers (which are not included in our sample) and smooth accretion
from the IGM.

\end{document}